\documentclass[manuscript]{acmart}

\usepackage{booktabs, multirow} 
\usepackage{soul}
\usepackage{pifont}
\usepackage{xspace}
\usepackage{tabularx}
\usepackage{colortbl}
\usepackage{float}
\usepackage{subcaption,booktabs}
\usepackage{siunitx}
\usepackage{tablefootnote}
\usepackage{array, makecell}
\usepackage{macros}

\newcommand{\cmark}{\ding{51}}%
\newcommand{\xmark}{\ding{55}}%

\AtBeginDocument{%
  \providecommand\BibTeX{{%
    \normalfont B\kern-0.5em{\scshape i\kern-0.25em b}\kern-0.8em\TeX}}}

\copyrightyear{2022}
\acmYear{2022}
\setcopyright{acmcopyright}\acmConference[UMAP '22]{Proceedings of the 30th ACM
Conference on User Modeling, Adaptation and Personalization}{July 4--7,
2022}{Barcelona, Spain}
\acmBooktitle{Proceedings of the 30th ACM Conference on User Modeling, Adaptation
and Personalization (UMAP '22), July 4--7, 2022, Barcelona, Spain}
\acmPrice{15.00}
\acmDOI{10.1145/3503252.3531292}
\acmISBN{978-1-4503-9207-5/22/07}

\usepackage{todonotes}
\presetkeys
    {todonotes}
    {inline,backgroundcolor=yellow}{}

\usepackage{xspace}
\usepackage{multirow}

\usepackage[subtle]{savetrees}

\begin{document}

\title{Top-N Recommendation Algorithms: A Quest for the State-of-the-Art}


\settopmatter{printacmref=false}

\author{Vito Walter Anelli}
\email{vitowalter.anelli@poliba.it}
\affiliation{\institution{Politecnico di Bari}
    \country{Italy}
 }

\author{Alejandro Bellogín}
\email{alejandro.bellogin@uam.es}
 \affiliation{
 \institution{Universidad Autónoma de Madrid}
 \country{Spain}
 }

\author{Tommaso Di Noia}
\email{tommaso.dinoia@poliba.it}
\affiliation{\institution{Politecnico di Bari}
\country{Italy}
 }
 
\author{Dietmar Jannach}
\email{dietmar.jannach@aau.at}
\affiliation{\institution{University of Klagenfurt}
\country{Austria}
}

\author{Claudio Pomo}
\email{claudio.pomo@poliba.it}
\affiliation{\institution{Politecnico di Bari}
\country{Italy}
 }

\renewcommand{\shortauthors}{Anelli et al.}

\begin{abstract}
Research on recommender systems algorithms, like other areas of applied machine learning, is largely dominated by efforts to \emph{improve the state-of-the-art}, typically in terms of accuracy measures. Several recent research works however indicate that the reported improvements over the years sometimes ``don't add up'', and that methods that were published several years ago often outperform the latest models when evaluated independently. Different factors contribute to this phenomenon, including that some researchers probably often only fine-tune their own models but not the baselines.

In this paper, we report the outcomes of an in-depth, systematic, and reproducible comparison of ten collaborative filtering algorithms---covering both traditional and neural models---on several common performance measures on three datasets which are
frequently used for evaluation in the recent literature. Our results show that there is no consistent winner across datasets and metrics for the examined \emph{top-n} recommendation task. 
Moreover, we find that for none of the accuracy measurements any of the considered neural models led to the best performance. Regarding the performance ranking of algorithms across the measurements, we found that linear models, nearest-neighbor methods, and traditional matrix factorization consistently perform well for the evaluated modest-sized, but commonly-used datasets. Our work shall therefore serve as a guideline for researchers regarding existing baselines to consider in future performance comparisons. Moreover, by providing a set of fine-tuned baseline models for different datasets, we hope that our work helps to establish a common understanding of the state-of-the-art for \emph{top-n} recommendation tasks.
\end{abstract}

\ccsdesc[500]{Information systems~Recommender systems}
\keywords{Recommender Systems, Performance Comparison, Reproducibility}

\maketitle


\section{Introduction}
Recommender systems are nowadays widely used in online applications, where they help users find relevant information in situations of information overload. Given the high practical relevance of such systems, research in this field is flourishing, particularly in the underlying machine learning (ML) algorithms used to create personalized item suggestions. Correspondingly, the predominant methodology is offline experimentation where the prediction or ranking accuracy of different ML models is compared. The common goal in such research works is to advance the \emph{state-of-the-art}, and evidence is then provided by reporting improvements over existing models that were obtained in those experiments.

Unfortunately, a number of recent research works published in the area of recommender systems and other related areas of applied ML research, e.g., information retrieval, indicate that some of these improvements that have been reported over the years ``don't add up'' \citep{Armstrong:2009:IDA:1645953.1646031}.
\citet{Ferraridacrema2020troubling}, for example, benchmark a variety of recent \emph{top-n} recommendation models against earlier and often simpler models. Through their studies, they found that much of the reported progress only seems to be ``virtual'', as the latest models are almost always outperformed by existing methods (see also \citet{ncfvsmf2020} for a related analysis). Various reasons may contribute to this surprising phenomenon, including the choice of (too weak) baselines \cite{latifi2021ins,lin2019neural} or the lack of a proper tuning of the baselines. Moreover, in such independent evaluations, i.e., that are not done by authors of the compared methods, it often turns out that there is no clear winner across datasets and accuracy measures. Thus, it remains unclear what represents the actual state-of-the-art in this field, given that the ranking of algorithms seems to depend on the particular experimental configuration in terms of baselines, accuracy measures, or datasets.

With this work, our goal is to provide insights regarding what represents the state-of-the-art for \emph{top-n} recommendation tasks, at least for those experimental settings that are common in the recent literature. Like in \citet{Ferraridacrema2020troubling}, we consider a broad range of collaborative filtering algorithms, which includes both older methods based on nearest-neighbors, different matrix factorization approaches, linear models, as well as more recent techniques based on deep learning. Differently from earlier comparisons like \citet{Ferraridacrema2020troubling}, however, we benchmark all algorithms under identical experimental conditions, i.e., with the same datasets and using the same evaluation protocol, after systematically tuning the hyperparameters of all models to reach their best performance.\footnote{We share all code and data used to run the experiments publicly to ensure reproducibility of our findings, see 
\underline{\href{https://github.com/sisinflab/Top-N-Recommendation-Algorithms-A-Quest-for-the-State-of-the-Art}{our GitHub repository}.
}}

The outcomes of our experiments show that in none of the considered cases one of the two recent neural methods was the best-performing algorithm. Moreover, the ranking of the algorithms, as expected from the literature, varies across datasets and evaluation measures. 
With some surprise, we found that linear models, nearest neighbors, and traditional matrix factorization and are dominating the leaderboard across datasets and performance metrics. One insight from our research therefore is that these top-ranking non-neural methods from our analysis should be considered as baselines in future research on recommendation algorithms. 

It is worth noticing that the datasets used in our experiments were chosen based on the predominant practice in the current academic literature. In our view, these datasets are however relatively small and different results might be obtained for larger datasets. Such an analysis is however not the focus of our present work, which aims to provide insights on the state-of-the-art in commonly used evaluation setups. Nonetheless, with this work we provide a set of fine-tuned models for these common datasets, thereby reducing the effort for other researchers to tune these baselines in their own experiments. In the future, we plan to publish fine-tuned models also for larger datasets, thereby continuously growing our understanding of the state-of-the-art in this area.

The paper is organized as follows. Next, in Section~\ref{sec:methodology}, we describe the details of our methodology and the datasets, algorithms, and metrics that we used in our experiments. Section \ref{sec:results} discusses the outcomes of our experiments, both in terms of accuracy and beyond-accuracy metrics. Section~\ref{sec:summary} finally discusses and summarizes the insights of our research and provides an outlook on future works. \label{sec:intro}

\color{black}
\section{Methodology}\label{sec:methodology}
The goal of our study was to evaluate different algorithms under very \emph{common} experimental settings in the current literature in terms of datasets, evaluation metrics, and protocols. The choice of experimental settings reported in this paper were guided by the following considerations.
First, we took inspiration from the work by~\citet{SunAreWeEvaluating2020}, who systematically evaluated various algorithms under a large set of experimental configurations. Second, to select specific experimental configurations for the purpose of our study, we scanned the current literature for the rather common settings. This also led to the inclusion of a number of recent models as well as simpler methods that have proven effective in recent works, where some of them had not been considered in~\citeauthor{SunAreWeEvaluating2020}.

Notably, our present work generally differs from~\citeauthor{SunAreWeEvaluating2020} in terms of the main goal. In \citeauthor{SunAreWeEvaluating2020}, one main purpose was to assess the impact of various aspects of the experimental procedure, e.g., negative sampling, split-ratio, or dataset preprocessing, 
on accuracy. In contrast, our work mainly focuses on providing a performance comparison of algorithms of different families for very common experimental configurations. Thus, we hope that our work helps establish an agreed-upon and continuously updated benchmark setting that can be used for researchers to test their new models against existing ones in a predefined setting.\footnote{We note that some of the choices regarding the experimental settings could have been made differently as well, e.g., with respect to cutoff thresholds. We however do not expect largely different results when changing some of these minor experiment parameters.}

\subsection{Datasets and Preprocessing}
\label{subsec:datasets}
We report the results we obtained for three datasets that are frequently used in the recent literature: \emph{MovieLens-1M}, \emph{Amazon Digital Music}, and \emph{Epinions}.

\begin{itemize}
  \item \emph{MovieLens-1M} (\ml): The MovieLens datasets have been widely used in the recommender systems literature for many years~\cite{HarperML2015} and different versions are available online\footnote{\url{https://grouplens.org/datasets/movielens/}}. The \ml dataset used in our studies was collected between the years 2000 and 2003 on the MovieLens website and contains ratings for movies on a 1-5 scale. A particularity of the dataset is that it is rather dense, and for each user at least 20 ratings are available.
  \item \emph{Amazon Digital Music} (\amzmusic): This dataset is part of a larger public collection of datasets\footnote{\url{https://jmcauley.ucsd.edu/data/amazon/}} that was created initially in the context of image-based recommendation~\cite{McAuley2015Image}. The Digital Music dataset contains reviews crawled from the Amazon website as well as item ratings on a 1-5 scale.
  \item \emph{Epinions}: This dataset was crawled in 2003 from the now defunct consumer review site \texttt{epinions.com}\footnote{\url{http://www.trustlet.org/epinions.html}}.
  A peculiar characteristic of the Epinions website was that users were paid according to how much a review was found useful. For this reason, Epinions has been widely adopted for research on \emph{trust} in recommender systems. The Epinions collection consists of two datasets: one contains item ratings (1-5 stars), while the other one collects (unary) trust statements among users. We point out that, in our study, instead of setting a custom (and in some ways arbitrary) threshold to binarize the rating dataset, we use the second dataset and consider the ``trustable'' users as the objective of the recommendation task.
\end{itemize}

It is noteworthy that, for the purpose of our research, all three datasets are publicly available and they were selected also in order to cover a diverse set of application domains of recommender systems. Other datasets, e.g., from the Netflix Prize, were also popular for some time, but they are nowadays only rarely used, e.g.,~\citet{DBLP:conf/www/LiangKHJ18}, and they are no longer officially accessible.
Moreover, differently from the Netflix Prize competition, \emph{rating prediction} is also no longer considered the most important task in recommendation. Instead, the common goal nowadays is to compute \emph{item rankings}. In addition, recommending based on implicit feedback signals is dominating the landscape, given the typical lack of explicit rating information in many applications. Therefore, datasets that originally contain item ratings 
are commonly converted into unary (like) signals. We follow this practice also in our evaluation and convert the rating datasets of MovieLens and Amazon Digital Music to unary datasets  by considering every rating above 3 as a positive signal.\footnote{Alternative approaches exist in the literature for this conversion, e.g., considering every rating as positive in case it is higher than the user's average. Often, we also see that all ratings are converted to positive signals. This is however questionable as \emph{(i)} a low rating, e.g., one star, is not a positive signal and \emph{(ii)} it changes the problem into predicting who will rate what.} For the Epinions dataset, such a conversion is not needed as the data is already given in unary form.

Real-world datasets are often very sparse. Therefore, another common pre-processing step in the literature is to create a more dense version of the datasets to ensure that there is a minimum number of interactions per user and item in the dataset, e.g., to allow for effective personalization.
We created different \emph{p-cores} for each dataset due to their diverging characteristics. In a \emph{p-core} dataset, we ensure that there are at least $p$ interactions for each item and 
at least $p$ interactions for each user. For our experiments, the creation of these $p$-core datasets was done in an iterative procedure, where the described constraints are applied until no more changes to the dataset can be observed. Different values for $p$ were used for the given datasets, depending on their size and density. For Movielens 1M and Amazon Digital Music we used the most common value of $p$ ($p$=10 for ML1M, $p$=5 for \amzmusic, see~\citet{SunAreWeEvaluating2020}), while for Epinions we choose a $p$-core value to reach a comparable density of the final matrix with respect to the other two datasets. Specifically, in this latter case, only a 2-core subset was computed due to the dataset's high sparsity. The resulting dataset characteristics are shown in Table~\ref{tab:datasets}. We observe that removing negative ratings and creating p-cores led to a considerable reduction of the dataset size for \ml, and that it results in an even more drastic reduction for the \amzmusic dataset.

\begin{table*}[h!t]
\caption{Dataset characteristics before and after pre-processing}
\label{tab:datasets}
\rowcolors{3}{gray!15}{white}
\begin{tabular}{lrrrrrrr}
  \hline
  \textbf{Dataset} & \textbf{p-core}  & \textbf{\#interactions} & \textbf{\#users} & \textbf{\#items} & \textbf{\#interactions} & \textbf{\#users} & \textbf{\#items} \\ \hline
  \multicolumn{2}{l}{} & \multicolumn{3}{c}{\textit{before pre-processing}} &  \multicolumn{3}{c}{\textit{after pre-processing}} \\ \hline
  Movielens 1M & 10-core & 1,000,209 & 6,040 & 3,706 & 571,531 & 5,949 & 2,810 \\
  Amazon Digital Music & 5-core & 1,584,082 & 840,372 & 456,992 & 145,523 & 14,354 & 10,027 \\
  Epinions & 2-core &   300,548 & 8,514 & 8,510\textsuperscript{\textdagger} & 300,475 & 8,485 & 8,463\textsuperscript{\textdagger} \\
  \hline
\end{tabular}
\caption*{\footnotesize \textdagger: The Epinions dataset focuses on ``trustable'' user recommendation. Note that not all users are trustable candidates according to the historical transactions, which is why the number of users as recommendable items is lower than the number of users.}
\end{table*}

Interestingly, today's commonly used datasets are often not only almost twenty years old, but also rather small, compared, for example, to the Netflix Prize dataset with its 100 million ratings. We assume that the computational complexity of some modern  models prevents authors to explore their proposals on larger datasets. Among the larger \emph{public} datasets, the 20M version of the MovieLens datasets is sometimes used in the literature~\citep{DBLP:conf/www/LiangKHJ18}. The 1M version is however used for evaluations more frequently~\cite{SunAreWeEvaluating2020}, and this is the main reason why we consider it in this study. Moreover, we observed that systematically tuning the hyperparameters for all datasets and models can be computationally challenging for some models already for the datasets of modest size described in Table~\ref{tab:datasets}.

\subsection{Algorithms}
\label{subsec:algorithms}
Given the goals described above, we considered algorithms from different families in our analysis. All non-neural methods, except Bayesian Personalized Ranking (BPRMF)~\cite{DBLP:conf/uai/RendleFGS09}, were also considered as baselines in the recent analysis of recommendation algorithms presented in~\citet{Ferraridacrema2020troubling}. Specifically, we considered the following techniques in our evaluation:

\begin{itemize}
  \item Non-personalized baseline: Popularity-based recommendation (\emph{MostPop}).
  \item Neighborhood-based and simple graph-based models: \userknn \cite{ResnickGrouplens1994}, \itemknn \cite{sarwas2001itembsed}, \pbeta \cite{DBLP:journals/tiis/PaudelCNB17}.
  \item Linear models: \slim \cite{DBLP:conf/icdm/NingK11}, \easer \cite{DBLP:conf/www/Steck19}.
  \item Matrix factorization models: BPRMF \cite{DBLP:conf/uai/RendleFGS09}, MF2020 \cite{ncfvsmf2020}, \ials \cite{DBLP:conf/icdm/HuKV08}.
  \item Neural models: NeuMF \cite{he2017neural}, 
  MultiVAE \cite{DBLP:conf/www/LiangKHJ18}.
\end{itemize}

Table~\ref{tab:algorithms} provides more details for the compared algorithms and explains why we considered them for our study.
\begin{table*}[h!t]
\caption{Overview of compared algorithms}
\label{tab:algorithms}
\begin{tabular}{p{3cm}lp{9cm}}
  \hline
  \textbf{Family} & \textbf{Algorithm} & \textbf{Description} \\ \hline
  \multirow{2}{3cm}{Non-personalized Baselines} & MostPop & Recommends the most popular items to each user, where popularity is defined by the number of observed interactions in the training data.\\ \cline{2-3}
  & Random & Creates random recommendations for users. Mainly useful to provide a reference point for beyond-accuracy measures (see Section~\ref{subsec:evaluation-setup}). \\ \hline
  \multirow{3}{3cm}{Neighbors and Graphs} & \userknn & A user-based nearest neighbor scheme proposed by \citet{ResnickGrouplens1994} in 1994 in an early  paper on the GroupLens system. In general, we include early nearest-neighbor techniques here because \emph{(i)} they let us gauge the progress on small datasets over time and \emph{(ii)} they proved surprisingly effective in recent research \cite{Ferraridacrema2020troubling}.\\ \cline{2-3}
   & \itemknn & Item-based nearest-neighbor algorithms were discussed in 2001 \cite{sarwas2001itembsed} and later successfully applied in industry around 2003 \cite{DBLP:journals/internet/LindenSY03}.  \\ \cline{2-3}
  & \pbeta & This method (\pbeta) is a simple graph-based method \cite{DBLP:journals/tiis/PaudelCNB17} from 2017, which is conceptually similar to the \itemknn method and can, despite its simplicity, lead to good performance \citep{Ferraridacrema2020troubling}. \\ \hline
  \multirow{2}{3cm}{Linear Models} & \slim & This regression-based method was proposed for top-\emph{n} recommendation tasks in 2011 \cite{DBLP:conf/icdm/NingK11}. Like in a recent analysis \cite{Ferraridacrema2020troubling}, we use the \emph{ElasticNet} version of the method \cite{levy2013SLIM_ElasticNet}, as it often leads to competitive results. \\ \cline{2-3}
   & \easer & Another linear model, proposed in 2019 \cite{DBLP:conf/www/Steck19}, which works like a shallow autoencoder. We include this method because it often leads to good results despite its simplicity. \\ \hline
  \multirow{3}{3cm}{Matrix Factorization}    & MF2020 & Matrix factorization methods were initially explored using Singular Value Decomposition in 1998 \cite{Billsus1998}. Later, in particular during and after the Netflix Prize, various machine learning approaches were proposed to learn latent factors\footnotemark. A recent analysis shows that these methods from the late 2000s are still competitive. In our study, we use a very recent MF model from \citet{ncfvsmf2020} proposed in 2020, dubbed MF2020. \\ \cline{2-3}
  & \ials & This method from 2008 uses an Alternating Least Squares approach and is particularly designed to learn factor models for implicit feedback datasets \cite{DBLP:conf/icdm/HuKV08}. The method is widely used as a non-neural baseline in the literature. \\ \cline{2-3}
  &  BPRMF & This method from 2009 was also designed for implicit feedback and introduces a novel optimization criterion. We use the MF variant in our experiments, which is also frequently used as a non-neural baseline in the literature~\cite{DBLP:conf/uai/RendleFGS09}. \\ \hline
  \multirow{2}{3cm}{Neural Models} & NeuMF   & NeuMF was proposed in 2017 \cite{he2017neural} and is an early and influential deep learning model used for recommendation. It generalizes matrix factorization and replaces the inner product with a neural architecture. The method is widely used as a neural baseline in the recent literature.\\ \cline{2-3}
   & MultiVAE & This model was designed for implicit feedback data, published in 2018, and is based on variational autoencoders \cite{DBLP:conf/www/LiangKHJ18}. According to the analysis in \citet{Ferraridacrema2020troubling}, this method outperformed existing non-neural baselines in an independent evaluation.\\ \hline
\end{tabular}
\end{table*}
\footnotetext{\url{https://sifter.org/simon/journal/20061211.html}}

\subsection{Evaluation Settings and Metrics}
\label{subsec:evaluation-setup}
In this section, we provide details about the applied evaluation protocol, the evaluation metrics, and the hyperparameter tuning process.

\paragraph{Evaluation Protocol}
We used a common \emph{repeated} 80-20 hold-out splitting procedure in our experiments~\cite{ncfvsmf2020}. Correspondingly, each dataset is randomly split to sample chunks containing around 20\% of the data. In each evaluation round, 20\% of the data are used for testing and the remaining 80\% are for training. Each experiment is repeated five times.
Later in Section~\ref{sec:results}, we report the mean of the observed values of the cross-validation runs.

We note that in the recent literature often only the results of one single training-test split are reported. While this 
data-splitting is typically done randomly in previous studies, we argue that cross-validation usually leads to more reliable results.

\paragraph{Metrics}
We collect a rich variety of accuracy metrics as well as a number of ``beyond-accuracy'' measures that are commonly used in the literature to assess additional quality aspects of recommendation lists.

\begin{itemize}
\item
In terms of \emph{accuracy metrics}, we measure Normalized Discounted Cumulative Gain (nDCG), Mean Reciprocal Rank (MRR), Precision, Recall,  Mean Average Precision (MAP), and F1 at common list lengths of 10, and 20.
For F1, note that we compute it on a per-user basis and not simply as a harmonic mean of the averages of Precision and Recall across users.\footnote{With this user-wise calculation of F1, the  overall average of F1 values is not bounded to lie between the overall averages of Precision and Recall; see the online material for additional explanations (\url{https://github.com/sisinflab/Top-N-Recommendation-Algorithms-A-Quest-for-the-State-of-the-Art})}
Thus, we have both metrics that take the position of the correct items into account and metrics that are agnostic of this aspect. Note here that we do not collect ``sampled'' metrics in our evaluation. In a \emph{sampled metrics} approach, one test item is ranked within an often small list of ``negative samples''. Such a procedure, while widely used, was recently found to be unreliable~\cite{KricheneSampledMetrics2020}. Note that historically the majority of the literature considered error metrics (RMSE, MAE) for evaluation purposes. However,  ``\emph{such classical error criteria do not really measure top-N performance}''~\cite{DBLP:conf/recsys/CremonesiKT10}. 
Consequently, several ranking metrics have been proposed in the last two decades and were adopted to evaluate top-n recommendation tasks. The present work shows the evaluation results for the most commonly used ranking metrics.

\item Considering \emph{beyond-accuracy metrics}, we measured a broader range of metrics regarding popularity bias, novelty, fairness, and item coverage and concentration. The details of the considered metrics are provided in Table~\ref{tab:metrics}. We note that also the novelty and fairness metrics used here are based on popularity distributions of items.  Specifically, for the PRSP and the PREO metrics, we consider the 20\% most popular items as the ``short head'' and the rest as long-tail items.
\item \emph{Running times:} Modern machine learning models can be computationally expensive. Therefore, we measured the computation times required for each algorithm for training and testing.
\end{itemize}

\begin{table*}[h!t]
\caption{Overview of beyond-accuracy metrics}\vspace{-1em}
\label{tab:metrics}
\begin{tabular}{p{3cm}lp{9cm}}
  \hline
  \textbf{Aspect} & \textbf{Metric} & \textbf{Description} \\ \hline
  \multirow{3}{3cm}{Coverage and Concentration} & \textbf{IC} & Item Coverage (IC) measures how many items ever appear in the top-\emph{n} recommendations of users.\\ \cline{2-3}
   & \textbf{Gini} & A measure of statistical dispersion, used to express the inequality of a distribution. A higher Gini index value ($\mathtt{Gini} \in [0,\ldots,1]$) indicates a stronger concentration of the recommendations, e.g., on popular items \cite{JannachLercheEtAl2015}.
   To ease the interpretation of the results and associate higher values with better results in terms of non-concentrated recommendations, in Tables  \ref{tab:beyond-accuracy-ml1m}, \ref{tab:beyond-accuracy-amzm}, and \ref{tab:epinions_beyond} we report the value $(1-\texttt{Gini})$.\\ \hline
     \multirow{2}{3cm}{Novelty} & \textbf{EFD} & Expected Free Discovery: A novelty measure proposed in \cite{vargas2011} based on the \emph{inverse collection frequency}. Like EPC, this measure expresses the ability of an algorithm to recommend relevant long-tail items. \\ \cline{2-3}
     & \textbf{EPC} & Expected Popularity Complement: This metric expresses the \emph{expected ``number of seen items not previously seen''}~\cite{vargas2011}. \\ 
    \hline
   \multirow{3}{3cm}{Fairness}    & \textbf{PREO} & The Popularity-based Ranking-based Equal Opportunity (REO) recommendation metric for assessing bias (fairness) was proposed in \cite{Zhu2020measuring}. Lower values mean less biased recommendations. \\ \cline{2-3}
  & \textbf{PRSP}  & Popularity-based Ranking-based Statistical Parity \cite{Zhu2020measuring}, to assess potential bias and thus fairness of the recommendations. Again, lower values mean less biased recommendations. \\  \hline
  \multirow{2}{3cm}{Popularity Bias} & \textbf{APLT} & Average Popularity of Long-Tail Items: Measures the average popularity of long tail items in the top-\emph{n} recommendations of users \cite{abdollahpouri2019managing}. \\ \cline{2-3}
  & \textbf{ARP} & Average Rating-based Popularity: This metric computes the popularity of the items in a recommendation list based on the number of interactions of each item in the training data \cite{JannachLercheEtAl2015}. \\ \cline{2-3}
  & \textbf{ACLT} &Average Coverage of Long-Tail Items: Measures how many items from the long tail are covered in the top-\emph{n} recommendations of users \cite{abdollahpouri2019managing}. \\
   \hline
\end{tabular}\vspace{-1em}
\end{table*}

\paragraph{Hyperparameter tuning}
We performed extensive hyperparameter tuning for all algorithms in our comparison, which is essential to understand what represents the \emph{state-of-the-art}. Previous research~\cite{cremonesi2021aimag} has identified that the lack of proper tuning of baseline algorithms may easily lead to a certain stagnation in the field, where new models are carefully tuned, whereas only limited effort sometimes goes into tuning existing baseline models.

For hyperparameter tuning, we relied on the HyperOpt library\footnote{\url{http://hyperopt.github.io/hyperopt/}} and used Tree of Parzen Estimators (TPE) as an algorithm to find the best hyperparameters~\cite{Bergstra2011}.  We determined suitable hyperparameter ranges for each algorithm from the literature, using, e.g., ranges that were earlier used in~\citet{Ferraridacrema2020troubling} and other works. Depending on the number and ranges of the hyperparameters of each algorithms, we explored between 20 and 50 hyperparameter combinations for each model. Hyperparameter tuning was conducted on a validation set for each dataset, and nDCG@10 was used as an optimization target.
As suggested by \citet{DBLP:conf/recsys/AnelliNSPR19}, the nDCG metric represents a reasonable choice for hyperparameter tuning.
All hyperparameter ranges and the optimal values for each dataset and algorithm are reported in the provided online material for reproducibility.


\section{Results}\label{sec:results}


\subsection{Accuracy Results}
\definecolor{Gray}{gray}{0.9}
\newcolumntype{g}{>{\columncolor{Gray}}c}

The results of the accuracy measurements for commonly used cutoff thresholds of 10 and 20 are shown in Table~\ref{tab:movielens_accuracy} (MovieLens-1M), Table~\ref{tab:amzon_dm_accuracy} (Amazon Digital Music), and Table~\ref{tab:epinions_accuracy} (Epinions). The results for the cutoff threshold of 50 are provided in the online material. 
We mark the best-performing method for each metric in bold font; the second-best result is underlined. The following main observations can be made.

\begin{itemize}
  \item \emph{Top-performing methods:} Considering nDCG as our main performance measure---most other metrics are correlated except for Recall in some situations---we find that the top three positions across all metrics and cutoff lengths are taken by five algorithms: \easer, MF2020, \slim, \pbeta, and, a bit surprisingly, \userknn. Differences across the datasets exist, but the ranking at least at top places is quite consistent across the datasets. For \ml, \easer, MF2020, and \slim are the best methods, whereas \pbeta, \easer, and \slim are best for \amzmusic. These methods also work well in Epinions. For the Epinions dataset, however, 
  \userknn works even slightly better than \easer. Generally, the performance of the five top-performing methods is quite consistent, \emph{with \easer always taking a top rank}. The MF2020 technique, in contrast, mainly seems to work particularly well for the dense ML1M dataset.
  We note here that \userknn for the given datasets was always favorable over \itemknn. It is noticeable that this evidence differs from some prior literature.
In 2004~\cite{DBLP:journals/tois/DeshpandeK04}, it was suggested
that item-based algorithms provide comparable or better quality recommendations than traditional user-neighborhood-based recommender systems. In 2011, researchers reported~\cite{DBLP:conf/icdm/NingK11} that in their experiments item-based schemes outperform user-based ones. Similar observations were made in 2011 by \citet{DBLP:conf/recsys/EkstrandLKR11} for rating prediction tasks.
In 2016,~\citet{DBLP:conf/recsys/Christakopoulou16} generally assumed that the item-based methods had been shown to outperform the user-based schemes for the \emph{top-n} recommendation task. In the analysis from 2021~\cite{Ferraridacrema2020troubling}, however, a general dominance of \itemknn over \userknn was not reported. There were cases where \itemknn was better, but in the majority of the reported experiments \userknn was favorable, which suggests that the ranking of the methods may depend on dataset characteristics and specifics of the evaluation protocol.

\begin{table*}[h]
\caption{Accuracy Results for MovieLens-1M. The tables are sorted by nDCG in descending order. The notation \emph{@N} indicates that the metrics are computed considering recommendation lists of \emph{N} elements.}\vspace{-1em}
\label{tab:movielens_accuracy}
\begin{minipage}{.49\linewidth}
\centering
\rowcolors{1}{white}{gray!15}
\renewcommand{\arraystretch}{0.8}
\footnotesize
\begin{tabular}{lrrrrrr}\hline
    \multirow{2}{*}{\textbf{Algorithm}} &\multicolumn{6}{c}{\textbf{Top@10}} \\
    \cmidrule(lr){2-7}
&\textbf{nDCG} &\textbf{MAP} &\textbf{MRR} &\textbf{Pre} &\textbf{Rec} &\textbf{F1}\\\midrule
\easer &\textbf{0.336} &\ul{0.335} &\textbf{0.583} &\ul{0.274} &\textbf{0.194} &\ul{0.190} \\
\slim &\ul{0.335} &\textbf{0.337} &\ul{0.580} &\textbf{0.275} &0.189 &0.188 \\
MF2020 &0.329 &0.327 &0.563 &0.272 &\ul{0.190} &\textbf{0.192} \\
\userknn &0.315 &0.314 &0.554 &0.256 &0.183 &0.179 \\
\pbeta &0.315 &0.313 &0.556 &0.256 &0.184 &0.179 \\
\ials &0.306 &0.304 &0.542 &0.252 &0.179 &0.176 \\
MultiVAE &0.294 &0.284 &0.514 &0.243 &0.183 &0.175 \\
\itemknn &0.292 &0.293 &0.518 &0.242 &0.163 &0.163 \\
NeuMF &0.277 &0.275 &0.494 &0.232 &0.157 &0.158 \\
BPRMF &0.275 &0.271 &0.502 &0.226 &0.166 &0.161 \\
MostPop &0.159 &0.159 &0.317 &0.137 &0.084 &0.086 \\
Random &0.008 &0.007 &0.020 &0.007 &0.004 &0.004 \\
\bottomrule
\end{tabular}
\end{minipage}
\begin{minipage}{.49\linewidth}
\centering
\rowcolors{2}{white}{gray!15}
\renewcommand{\arraystretch}{0.8}
\footnotesize
\begin{tabular}{lrrrrrr}\hline
    \multirow{2}{*}{\textbf{Algorithm}} &\multicolumn{6}{c}{\textbf{Top@20}} \\
    \cmidrule(lr){2-7}
&\textbf{nDCG} &\textbf{MAP} &\textbf{MRR} &\textbf{Pre} &\textbf{Rec} &\textbf{F1}\\\midrule
\easer &\textbf{0.335} &\ul{0.287} &\textbf{0.587} &\textbf{0.216} &\textbf{0.289} &\ul{0.206} \\
\slim &\ul{0.332} &\textbf{0.288} &\ul{0.584} &\textbf{0.216} &0.283 &0.204 \\
MF2020 &0.329 &0.283 &0.568 &\textbf{0.216} &\ul{0.286} &\textbf{0.207} \\
\pbeta &0.315 &0.269 &0.561 &0.203 &0.277 &0.195 \\
\userknn &0.314 &0.268 &0.559 &0.201 &0.273 &0.192 \\
\ials &0.309 &0.263 &0.547 &0.202 &0.272 &0.194 \\
MultiVAE &0.304 &0.250 &0.519 &0.199 &0.281 &0.195 \\
\itemknn &0.289 &0.252 &0.523 &0.192 &0.247 &0.180 \\
BPRMF &0.280 &0.235 &0.508 &0.181 &0.253 &0.176 \\
NeuMF &0.280 &0.240 &0.500 &0.188 &0.245 &0.195 \\
MostPop &0.161 &0.141 &0.326 &0.114 &0.137 &0.103 \\
Random &0.009 &0.007 &0.024 &0.007 &0.007 &0.006 \\
\bottomrule
\end{tabular}
\end{minipage}
\end{table*} 

\begin{table*}[h]
\caption{Accuracy Results for Amazon Digital Music. The tables are sorted by nDCG in descending order.}\vspace{-1em} 
\label{tab:amzon_dm_accuracy}
\begin{minipage}{.49\linewidth}
\centering
\rowcolors{1}{white}{gray!15}
\renewcommand{\arraystretch}{0.8}
\footnotesize
\begin{tabular}{lrrrrrr}\hline
    \multirow{2}{*}{\textbf{Algorithm}} &\multicolumn{6}{c}{\textbf{Top@10}}\\
    \cmidrule(lr){2-7}
&\textbf{nDCG} &\textbf{MAP} &\textbf{MRR} &\textbf{Pre} &\textbf{Rec} &\textbf{F1}\\\midrule
\pbeta &\textbf{0.085} &\textbf{0.040} &\textbf{0.115} &\textbf{0.023} &\ul{0.104} &\textbf{0.036} \\
\easer &\ul{0.083} &\ul{0.038} &\ul{0.108} &\textbf{0.023} &\textbf{0.106} &\ul{0.035} \\
\slim &0.081 &0.037 &0.106 &0.022 &\ul{0.104} &\ul{0.035} \\
\userknn &0.081 &0.037 &0.105 &0.022 &\ul{0.104} &\ul{0.035} \\
\ials &0.073 &0.032 &0.093 &0.021 &0.099 &0.033 \\
\itemknn &0.071 &0.033 &0.095 &0.018 &0.085 &0.029 \\
MF2020 &0.057 &0.024 &0.067 &0.017 &0.083 &0.028 \\
NeuMF &0.056 &0.024 &0.068 &0.015 &0.074 &0.025 \\
MultiVAE &0.054 &0.023 &0.062 &0.016 &0.077 &0.025 \\
BPRMF &0.020 &0.008 &0.023 &0.007 &0.031 &0.010 \\
MostPop &0.012 &0.005 &0.016 &0.004 &0.016 &0.006 \\
Random &0.000 &0.000 &0.000 &0.000 &0.001 &0.000 \\
\bottomrule
\end{tabular}
\end{minipage}
\begin{minipage}{.49\linewidth}
\centering
\rowcolors{2}{white}{gray!15}
\renewcommand{\arraystretch}{0.8}
\footnotesize
\begin{tabular}{lrrrrrr}\hline
    \multirow{2}{*}{\textbf{Algorithm}} &\multicolumn{6}{c}{\textbf{Top@20}} \\
    \cmidrule(lr){2-7}
&\textbf{nDCG} &\textbf{MAP} &\textbf{MRR} &\textbf{Pre} &\textbf{Rec} &\textbf{F1}\\\midrule
\pbeta &\textbf{0.094} &\textbf{0.029} &\textbf{0.118} &\textbf{0.015} &\ul{0.132} &\textbf{0.026} \\
\easer &\ul{0.092} &\ul{0.028} &\ul{0.111} &\textbf{0.015} &\textbf{0.136} &\textbf{0.026} \\
\slim &0.090 &0.027 &0.109 &0.014 &0.134 &0.025 \\
\userknn &0.090 &0.027 &0.108 &0.014 &0.133 &0.025 \\
\ials &0.084 &0.025 &0.096 &0.014 &0.132 &0.025 \\
\itemknn &0.078 &0.023 &0.098 &0.012 &0.108 &0.021 \\
MF2020 &0.067 &0.019 &0.071 &0.012 &0.116 &0.022 \\
NeuMF &0.063 &0.018 &0.071 &0.010 &0.096 &0.019 \\
MultiVAE &0.062 &0.017 &0.066 &0.011 &0.105 &0.019 \\
BPRMF &0.025 &0.007 &0.025 &0.005 &0.047 &0.009 \\
MostPop &0.014 &0.004 &0.017 &0.003 &0.025 &0.005 \\
Random &0.001 &0.000 &0.001 &0.000 &0.001 &0.000 \\
\bottomrule
\end{tabular}
\end{minipage}
\end{table*} 

\begin{table*}[h]
\caption{Accuracy Results for Epinions. The tables are sorted by nDCG in descending order. }\vspace{-1em}
\label{tab:epinions_accuracy}
\begin{minipage}{.49\linewidth}
\centering
\rowcolors{1}{white}{gray!15}
\renewcommand{\arraystretch}{0.8}
\footnotesize
\begin{tabular}{lrrrrrr}\hline
    \multirow{2}{*}{\textbf{Algorithm}} &\multicolumn{6}{c}{\textbf{Top@10}}\\
    \cmidrule(lr){2-7}
&\textbf{nDCG} &\textbf{MAP} &\textbf{MRR} &\textbf{Pre} &\textbf{Rec} &\textbf{F1}\\\midrule
\userknn &\textbf{0.164} &\ul{0.131} &\ul{0.266} &\ul{0.157} &\ul{0.102} &\textbf{0.100} \\
\easer &\textbf{0.164} &\textbf{0.132} &\textbf{0.268} &0.154 &\textbf{0.104} &\textbf{0.100} \\
\pbeta &0.163 &0.129 &0.260 &\textbf{0.159} &\ul{0.102} &\textbf{0.100} \\
\slim &0.156 &0.126 &0.254 &0.148 &0.101 &0.096 \\
MultiVAE &0.149 &0.116 &0.240 &0.148 &0.094 &0.094 \\
\itemknn &0.138 &0.111 &0.224 &0.133 &0.090 &0.087 \\
MF2020 &0.125 &0.104 &0.219 &0.114 &0.084 &0.079 \\
NeuMF &0.118 &0.098 &0.206 &0.108 &0.080 &0.075 \\
BPRMF &0.113 &0.093 &0.200 &0.106 &0.076 &0.071 \\
\ials &0.110 &0.091 &0.192 &0.101 &0.074 &0.084 \\
MostPop &0.045 &0.037 &0.093 &0.036 &0.029 &0.025 \\
Random &0.001 &0.001 &0.003 &0.001 &0.001 &0.001 \\
\bottomrule
\end{tabular}
\end{minipage}
\begin{minipage}{.49\linewidth}
\centering
\rowcolors{2}{white}{gray!15}
\renewcommand{\arraystretch}{0.8}
\footnotesize
\begin{tabular}{lrrrrrr}\hline
    \multirow{2}{*}{\textbf{Algorithm}} &\multicolumn{6}{c}{\textbf{Top@20}}\\
    \cmidrule(lr){2-7}
&\textbf{nDCG} &\textbf{MAP} &\textbf{MRR} &\textbf{Pre} &\textbf{Rec} &\textbf{F1}\\\midrule
\userknn &\textbf{0.178} &\ul{0.108} &\ul{0.272} &\ul{0.076} &\textbf{0.219} &\ul{0.093} \\
\easer &\ul{0.177} &\textbf{0.110} &\textbf{0.274} &\textbf{0.078} &0.216 &\textbf{0.094} \\
\pbeta &0.176 &0.107 &0.266 &0.075 &\ul{0.218} &0.092 \\
Slim &0.170 &0.106 &0.260 &0.076 &0.210 &0.091 \\
MultiVAE &0.165 &0.099 &0.247 &0.072 &0.212 &0.089 \\
\itemknn &0.151 &0.094 &0.230 &0.068 &0.190 &0.084 \\
MF2020 &0.138 &0.088 &0.225 &0.065 &0.170 &0.079 \\
NeuMF &0.131 &0.084 &0.212 &0.063 &0.162 &0.075 \\
BPRMF &0.126 &0.079 &0.207 &0.060 &0.160 &0.072 \\
\ials &0.121 &0.077 &0.198 &0.057 &0.149 &0.079 \\
MostPop &0.052 &0.032 &0.100 &0.025 &0.061 &0.029 \\
Random &0.002 &0.001 &0.003 &0.001 &0.002 &0.001 \\
\bottomrule
\end{tabular}
\end{minipage}
\end{table*} 


    \item \emph{Performance of neural methods:} The two neural methods considered here, NeuMF and MultiVAE, only led to medium performance on these datasets. While MultiVAE performed very well in an earlier comparison with traditional methods~\cite{Ferraridacrema2020troubling}, we may assume that the modest size of the datasets might limit the power of this method in our experiment to a certain extent, see also the report on the use of deep learning methods at Netflix~\cite{DBLP:journals/aim/SteckBELRB21} or the discussions in ~\citet{JannachdeSouzaetal2020}.
    \item \emph{Fine-tuning opportunities:} The \ials 
    and BPRMF methods often led to medium to modest performance in this comparison. Recent work indicates that further enhancing and fine-tuning methods like \ials for specific datasets may lead to additional performance improvements~\cite{Rendle2021iALS}. Note, however, that the goal of our work was to assess the performance of different algorithms under equal opportunities, i.e., by using a systematic but generic hyperparameter optimization procedure. Fine-tuning individual algorithms, e.g., by exploring rather uncommon ranges for the size of the latent factors, is of course possible, but not the focus of our work, which is about establishing a set of baselines (state-of-the-art) to consider in future works. Similar considerations apply for the neural methods, which may also be further tuned for individual datasets.

\end{itemize}

We note that the differences between the top-performing methods are sometimes small, often between one and a few percent. In papers that propose new models, we would therefore commonly expect statistical significance tests. For the evaluations reported in our study, we omit such tests as we have no prior hypotheses regarding which model would ``win''. Instead, the goal of our work is to provide guidance for researchers about which methods they might want to consider as baselines for comparison.
We note that in many published papers no exact details are provided about 
how the significance tests are applied and prerequisites were validated. Also, in case of per-user comparisons of means, significance at common $\alpha$-levels may be easy to achieve due to the large sample sizes~\cite{Lin2013TooBig}.

Comparing our algorithm ranking with earlier works~\cite{Ferraridacrema2020troubling,SunAreWeEvaluating2020}, we find both commonalities and differences. A general commonality of these studies is that more traditional methods, including linear models, matrix factorization, or nearest neighbors frequently take the top positions of the rankings. For example, the innovative combination of Factorization Machines with BPR loss worked particularly well in~\cite{SunAreWeEvaluating2020}. Also \slim and MF were in top positions for some datasets. Differently from our findings, NeuMF more often worked very well for some of the datasets examined in~\citet{SunAreWeEvaluating2020}. A competitive performance of NeuMF was also observed in~\citet{Ferraridacrema2020troubling}, where it was, however, usually slightly outperformed by various non-neural methods. These differences may be attributed to different causes, including specifics of data-preprocessing and the evaluation procedures\footnote{In the original paper proposing NeuMF, the authors for example used a \emph{leave-one-out} procedure where only the last item of each user was retained in the test set~\cite{he2017neural}.}. Differently from many earlier works, we
apply cross-validation and compute \emph{p-cores} iteratively instead of only filtering ``cold'' users and items once. Moreover, for some algorithms we explore a larger number of hyperparameter optimization trials than was done in some earlier works.

Finally, to obtain an overall picture of our accuracy results, we applied a Borda count \emph{ranked voting} scheme to aggregate the outcomes of our experiments. To that purpose, we consider each observed ranking for each dataset and metric as a vote. When applying the original Borda count scheme, each candidate (i.e., algorithm) receives more points when it is placed higher in the ranking. In our lists of 12 candidates, the first candidate receives 11 points and the last-ranked candidate 0 points. Applying this scheme across all accuracy measures at list length 10 leads to the ranking shown in Table~\ref{subtab:borda-count}.\footnote{The maximum possible value for a method in Table~\ref{subtab:borda-count} is 198, 
as we rank 12 algorithms according to 6 metrics for 3 datasets; 198=(12-1) $\times$ 1 $\times$ 3. For Table~\ref{subtab:borda-count-ndcg} and Table~~\ref{subtab:borda-count-recall}, the maximum is correspondingly 33. Although Tables \ref{tab:movielens_accuracy} to \ref{tab:epinions_accuracy} report rounded values for the sake of clarity, rankings are assessed considering exact metric values. 
} 


\begin{table*}[!htp]
\caption{Algorithm ranking based on Borda count at cutoff length 10.}
\begin{subtable}{0.33\textwidth}
\centering
\begin{table}[H]

\small
\rowcolors{2}{white}{gray!15}
\begin{tabular}{crc}\hline
  \textbf{Rank} & \textbf{Algorithm} & \textbf{Count} \\ \hline
1 & \easer & 185 \\
2 & \pbeta &  169 \\
3 & \slim &  160 \\
4 & \userknn &  154 \\
5 & MF2020 &  115  \\
6 & \itemknn &  99  \\
7 & MultiVAE &  92  \\
8 & \ials &  90  \\
9 & NeuMF &  61  \\
10 & BPRMF &  45  \\
11 & MostPop &  18  \\
12 & Random &  0  \\  \bottomrule
\end{tabular}
\caption{Overall}\label{subtab:borda-count}
\end{table}
\end{subtable}
\begin{subtable}{0.33\textwidth}
\centering
\begin{table}[H]
\small
\rowcolors{2}{white}{gray!15}
\begin{tabular}{crc}\hline
  \textbf{Rank} & \textbf{Algorithm} & \textbf{Count} \\ \hline
1 & \easer & 31 \\
2 & \userknn &  27 \\
3 & RP3beta &  27 \\
4 & \slim &  27 \\
5 & MF2020 &  19  \\
6 & \itemknn &  16  \\
7 &MultiVAE &  15  \\
8 & \ials &  13  \\
9 & NeuMF &  12  \\
10 & BPRMF &  7  \\
11 & MostPop &  3  \\
12 & Random &  0  \\  \bottomrule
\end{tabular}
\caption{nDCG}
\label{subtab:borda-count-ndcg}
\end{table}
\end{subtable}
\begin{subtable}{0.33\textwidth}
\centering
\begin{table}[H]
\small
\rowcolors{2}{white}{gray!15}
\begin{tabular}{crc}\hline
  \textbf{Rank} & \textbf{Algorithm} & \textbf{Count} \\ \hline
1 & \easer & 31 \\
2 & \pbeta &  29 \\
3 & \slim &  26 \\
4 & \userknn &  25 \\
5 & MF2020 &  20  \\
6 & MultiVAE &  17  \\
7 & \itemknn &  15  \\
8 & \ials &  14  \\
9 & NeuMF &  9  \\
10 & BPRMF &  9  \\
11 & MostPop &  3  \\
12 & Random &  0  \\  \bottomrule
\end{tabular}
\caption{Recall}
\label{subtab:borda-count-recall}
\end{table} 
\end{subtable}
\label{tab:borda_count_general}
\end{table*}

We emphasize that such a rank-based aggregation should be interpreted with great care as it might, for example, favor methods that work particularly well on a set of correlated metrics. 
%
In agreement with the analysis presented by \citet{DBLP:conf/recsys/ValcarceBPC18}, we observed high correlation between ranking metrics and for the same metric using different cutoffs.
For example, in that work, when computing the correlation between cutoffs ranging from $5$ to $100$, the lowest one was 0.9, which still represents a very strong correlation. Because of this, 
we only considered one threshold for the measurement shown in Table~\ref{subtab:borda-count}. 
Another known limitation of the Borda count scheme is that the ranking might change if a candidate is removed from the lists. Despite these limitations, we believe that the Borda count may represent a helpful summarization approach for the experiments in this paper.
More fine-grained applications of the Borda count are possible as well to account for such correlations. In Table~\ref{subtab:borda-count-ndcg} and Table~\ref{subtab:borda-count-recall}, we report the Borda count rankings when considering only one specific measure, nDCG@10 and Recall@10, respectively. We select Recall as an example here, because all other metrics are usually more correlated with nDCG than Recall. The analysis in Table~\ref{subtab:borda-count-recall} actually shows that \pbeta and \slim work particularly well for Recall and are ranked higher than \userknn for this metric.

\subsection{Beyond-Accuracy Results}
Table~\ref{tab:beyond-accuracy-ml1m} shows the beyond-accuracy metrics results for the MovieLens dataset for the top-10 and top-20 recommendations\footnote{Detailed results for other datasets and cutoff thresholds can be found in the online material.}. The rows in the table are again sorted by accuracy (nDCG).
We highlight the best values for each metric, not considering the Random and MostPop baselines, which only serve as reference points. Recommending random items will, for example, lead to high item coverage, but not to many relevant item suggestions.

In our analysis we found that some of our beyond-accuracy can be highly correlated, which is to some extent expected as many of them are based on item popularity characteristics, as discussed above.
Table~\ref{tab:correlations} shows the outcomes of an analysis of metric correlations. In this table, we report in how many cases (datasets) a metric is correlated with another one with a Pearson product-moment correlation coefficient (PPMCC) higher than 0.9 or lower than -0.9. We can observe that both the ACLT and the PRSP metrics are consistently correlated with the APLT metric. For the sake of conciseness, we therefore only report the APLT metric here and omit ACLT and PRSP from the tables. All detailed results also for these metrics can be found in the online material. 

\begin{table}[htbp]
\caption{Summary of Metric Correlations. A \cmark\ in a cell indicates a correlation of more than 0.9 (or beyond -0.9 vice versa) for one of the datasets. Two or three \cmark\ symbols mean that such a high correlation was also found for the second or the third dataset.}
\label{tab:correlations}
\rowcolors{1}{white}{gray!15}
\renewcommand{\arraystretch}{0.8}
\setlength\tabcolsep{0.3em}
\begin{tabular}{lcccccccc}\hline
\small
PPMCC & Gini & EFD & EPC & PREO & PRSP & ACLT & APLT & ARP \\ \hline
IC & \cmark & -- & -- & \cmark & -- & -- & -- & \cmark \\
Gini &  & -- & -- & \cmark & \cmark\cmark & \cmark & \cmark & -- \\
EFD &  &  & \cmark & -- & \cmark & \cmark & \cmark & -- \\
EPC &  &  &  & \cmark & -- & \cmark & \cmark & -- \\
PREO &  &  &  &  & \cmark & \cmark & \cmark & -- \\
PRSP &  &  &  &  &  & \cmark\cmark\cmark & \cmark\cmark\cmark & -- \\
ACLT &  &  &  &  &  &  & \cmark\cmark\cmark & -- \\
APLT &  &  &  &  &  &  &  & -- \\ \bottomrule
\end{tabular}
\label{tab:correlations}
\end{table}

\color{black}




\begin{table*}[!htp]\centering
\caption{Beyond Accuracy Results for MovieLens-1M. The tables are sorted by nDCG in descending order. The notation \emph{@N} indicates that the metrics are computed considering recommendation lists of \emph{N} elements. To ease the interpretation of the results and to associate higher values with more diversified recommendation lists, we report the value of $1-Gini$.}\label{tab:beyond-accuracy-ml1m}
\begin{minipage}{.49\linewidth}
\centering
\rowcolors{1}{}{gray!15}
\setlength{\tabcolsep}{0.315em}
\footnotesize
\begin{tabular}{lrrrrrrr}\hline
\multirow{2}{*}{\textbf{Algorithm}}&\multicolumn{7}{c}{\textbf{Top@10}}\\
    \cmidrule(lr){2-8}
&\textbf{IC} &\textbf{Gini} &\textbf{EFD} &\textbf{EPC} &\textbf{PREO} &\textbf{APLT} &\textbf{ARP}\\\midrule
\easer &838.0 &0.068 &\textbf{2.690} &\textbf{0.583} &0.978 &0.003 &1,062.727 \\
\slim &654.2 &0.052 &\underline{2.672} &0.244 &0.995 &0.001 &1,121.384 \\
MF2020 &920.2 &0.077 &\underline{2.672} &0.244 &0.968 &0.005 &1,042.373 \\
\userknn &1075.2 &0.067 &2.489 &\underline{0.227} &0.971 &0.010 &1,085.550 \\
\pbeta &854.4 &0.048 &2.461 &0.223 &0.959 &0.011 &1,181.638 \\
\ials &712.2 &0.080 &2.516 &0.232 &0.997 &0.000 &\underline{935.914} \\
MultiVAE &\textbf{1625.2} &\textbf{0.136} &2.422 &0.221 &\textbf{0.828} &\textbf{0.042} &\textbf{871.869} \\
\itemknn &1054.8 &0.066 &2.346 &0.214 &0.952 &0.011 &1,090.926 \\
NeuMF &\underline{1367.2} &\underline{0.111} &2.292 &0.209 &\underline{0.910} &\underline{0.028} &938.861 \\
BPRMF &1137.8 &0.091 &2.226 &0.203 &0.928 &0.010 &1,047.232 \\
MostPop &56.2 &0.005 &1.187 &0.103 &1.000 &0.000 &1,746.694 \\
Random &2810.0 &0.876 &0.074 &0.006 &0.039 &0.696 &151.045 \\
\bottomrule
\end{tabular}
\end{minipage}
\begin{minipage}{.49\linewidth}
\centering
\rowcolors{1}{}{gray!15}
\setlength{\tabcolsep}{0.315em}
\footnotesize
\begin{tabular}{lrrrrrrr}\hline
\multirow{2}{*}{\textbf{Algorithm}}
&\multicolumn{7}{c}{\textbf{Top@20}}\\
    \cmidrule(lr){2-8}
&\textbf{IC} &\textbf{Gini} &\textbf{EFD} &\textbf{EPC} &\textbf{PREO} &\textbf{APLT} &\textbf{ARP}\\\midrule
\easer &1093.0 &0.091 &\textbf{2.264} &\textbf{0.207} &0.963 &0.006 &949.860 \\
\slim &854.0 &0.069 &2.239 &0.205 &0.986 &0.003 &1,017.374 \\
MF2020 &1128.8 &0.095 &\underline{2.257} &\textbf{0.207} &0.946 &0.009 &969.260 \\
\pbeta &1207.2 &0.073 &2.085 &0.190 &0.937 &0.018 &1,024.408 \\
\userknn &1465.8 &0.092 &2.090 &0.191 &0.947 &0.017 &970.255 \\
\ials &901.000 &0.105 &2.138 &0.197 &0.983 &0.002 &\underline{838.660} \\
MultiVAE &\textbf{1924.8} &\textbf{0.156} &2.082 &0.190 &\textbf{0.792} &\textbf{0.052} &\textbf{821.849} \\
\itemknn &1346.0 &0.082 &1.977 &0.181 &0.939 &0.015 &1,006.978 \\
BPRMF &1386.2 &0.111 &1.893 &0.173 &0.890 &0.017 &968.784 \\
NeuMF &\underline{1679.4} &\underline{0.135} &1.965 &0.179 &\underline{0.867} &\underline{0.038} &868.452 \\
MostPop &92.0 &0.010 &1.039 &0.092 &1.000 &0.000 &1,570.672 \\
Random &2810.0 &0.911 &0.075 &0.007 &0.037 &0.693 &151.352 \\
\bottomrule
\end{tabular}
\end{minipage}
\end{table*} 
\begin{table*}[!htp]\centering
\caption{Beyond Accuracy Results for Amazon Digital Music. The tables are sorted by nDCG in descending order. }\label{tab:beyond-accuracy-amzm}
\begin{minipage}{.49\linewidth}
\centering
\rowcolors{1}{}{gray!15}
\setlength{\tabcolsep}{0.33em}
\footnotesize
\begin{tabular}{lrrrrrrr}\hline
\multirow{2}{*}{\textbf{Algorithm}}&\multicolumn{7}{c}{\textbf{Top@10}}\\
    \cmidrule(lr){2-8}
&\textbf{IC} &\textbf{Gini} &\textbf{EFD} &\textbf{EPC} &\textbf{PREO} &\textbf{APLT} &\textbf{ARP}\\\midrule
\pbeta &\textbf{9959.0} &\textbf{0.542} &\textbf{0.409} &\textbf{0.033} &\underline{0.308} &\underline{0.299} &\underline{23.759} \\
\easer &7789.0 &0.178 &\underline{0.368} &\underline{0.031} &0.537 &0.054 &56.155 \\
\slim &8215.4 &0.197 &0.361 &0.030 &0.552 &0.067 &49.287 \\
\userknn &7703.8 &0.181 &0.363 &0.030 &0.552 &0.056 &51.910 \\
\ials &4516.2 &0.136 &0.325 &0.027 &0.766 &0.009 &41.936 \\
\itemknn &\underline{9686.2} &\underline{0.478} &0.345 &0.027 &\textbf{0.097} &\textbf{0.550} &\textbf{9.884} \\
MF2020 &4722.8 &0.099 &0.242 &0.021 &0.687 &0.009 &59.949 \\
NeuMF &7365.2 &0.228 &0.245 &0.020 &0.455 &0.058 &30.236 \\
MultiVAE &6043.0 &0.189 &0.235 &0.020 &0.578 &0.045 &40.475 \\
BPRMF &3050.0 &0.024 &0.078 &0.007 &0.784 &0.001 &130.810 \\
MostPop &15.6 &0.001 &0.039 &0.004 &1.000 &0.000 &182.800 \\
Random &10025.8 &0.852 &0.002 &0.000 &0.229 &0.460 &9.840 \\
\bottomrule
\end{tabular}
\end{minipage}
\begin{minipage}{.49\linewidth}
\centering
\rowcolors{1}{}{gray!15}
\setlength{\tabcolsep}{0.33em}
\footnotesize
\begin{tabular}{lrrrrrrr}\hline
\multirow{2}{*}{\textbf{Algorithm}}
&\multicolumn{7}{c}{\textbf{Top@20}}\\
    \cmidrule(lr){2-8}
&\textbf{IC} &\textbf{Gini} &\textbf{EFD} &\textbf{EPC} &\textbf{PREO} &\textbf{APLT} &\textbf{ARP}\\\midrule
\pbeta &\textbf{10016.0} &\textbf{0.609} &\textbf{0.293} &\textbf{0.024} &\underline{0.318} &\underline{0.299} &\underline{22.353} \\
\easer &9441.0 &0.233 &\underline{0.267} &\underline{0.022} &0.542 &0.071 &47.987 \\
\slim &9659.4 &0.253 &0.263 &\underline{0.022} &0.548 &0.086 &43.337 \\
\userknn &9294.2 &0.237 &0.264 &\underline{0.022} &0.543 &0.073 &45.834 \\
\ials &5941.6 &0.177 &0.243 &0.020 &0.700 &0.015 &37.520 \\
\itemknn &\underline{9976.2} &\underline{0.528} &0.247 &0.019 &\textbf{0.121} &\textbf{0.546} &\textbf{9.870} \\
MF2020 &6389.4 &0.135 &0.187 &0.016 &0.661 &0.014 &51.479 \\
NeuMF &8533.2 &0.266 &0.180 &0.015 &0.468 &0.075 &27.416 \\
MultiVAE &9161.4 &0.329 &0.178 &0.015 &0.519 &0.094 &33.590 \\
BPRMF &4283.0 &0.034 &0.064 &0.006 &0.787 &0.002 &108.296 \\
MostPop &29.2 &0.002 &0.033 &0.004 &1.000 &0.000 &148.838 \\
Random &10025.8 &0.895 &0.002 &0.000 &0.158 &0.460 &9.838 \\
\bottomrule
\end{tabular}
\end{minipage}
\end{table*} 
\begin{table*}[!htp]\centering
\caption{Beyond Accuracy Results for Epinions. The tables are sorted by nDCG in descending order. }\label{tab:epinions_beyond}
\begin{minipage}{.49\linewidth}
\centering
\rowcolors{1}{}{gray!15}
\setlength{\tabcolsep}{0.33em}
\footnotesize
\begin{tabular}{lrrrrrrr}\hline
\multirow{2}{*}{\textbf{Algorithm}}
&\multicolumn{7}{c}{\textbf{Top@10}}\\
    \cmidrule(lr){2-8}
&\textbf{IC} &\textbf{Gini} &\textbf{EFD} &\textbf{EPC} &\textbf{PREO} &\textbf{APLT} &\textbf{ARP}\\\midrule
UserKNN &3402.2 &0.073 &\textbf{1.198} &\underline{0.112} &0.398 &0.080 &315.724 \\
EASER &2765.0 &0.055 &\underline{1.197} &\textbf{0.114} &0.501 &0.046 &341.486 \\
\pbeta &\textbf{6009.0} &\textbf{0.197} &1.237 &\underline{0.112} &\underline{0.198} &\textbf{0.275} &\underline{226.165} \\
\slim &3361.0 &0.081 &1.197 &0.110 &0.452 &0.061 &246.216 \\
MultiVAE &3386.8 &0.089 &1.105 &0.102 &0.364 &0.105 &293.641 \\
\itemknn &\underline{5832.8} &\underline{0.160} &1.084 &0.097 &\textbf{0.171} &\underline{0.267} &\textbf{214.681} \\
MF2020 &1235.2 &0.028 &0.921 &0.090 &0.786 &0.008 &368.003 \\
NeuMF &3595.0 &0.084 &0.905 &0.085 &0.495 &0.096 &322.865 \\
BPRMF &1322.8 &0.018 &0.824 &0.081 &0.651 &0.012 &475.629 \\
\ials &1613.4 &0.064 &0.891 &0.081 &0.707 &0.010 &214.989 \\
MostPop &41.6 &0.001 &0.273 &0.030 &1.000 &0.000 &719.301 \\
Random &8443.0 &0.823 &0.011 &0.001 &0.142 &0.738 &27.565 \\
\bottomrule
\end{tabular}
\end{minipage}
\begin{minipage}{.49\linewidth}
\centering
\rowcolors{1}{}{gray!15}
\setlength{\tabcolsep}{0.33em}
\footnotesize
\begin{tabular}{lrrrrrrr}\hline
\multirow{2}{*}{\textbf{Algorithm}}
&\multicolumn{7}{c}{\textbf{Top@20}}\\
    \cmidrule(lr){2-8}
&\textbf{IC} &\textbf{Gini} &\textbf{EFD} &\textbf{EPC} &\textbf{PREO} &\textbf{APLT} &\textbf{ARP}\\\midrule
UserKNN &4594.6 &0.095 &0.962 &\underline{0.090} &0.442 &0.083 &275.265 \\
EASER &3705.0 &0.071 &\textbf{0.965} &\textbf{0.091} &0.522 &0.047 &297.062 \\
\pbeta &\underline{7010.8} &\textbf{0.232} &0.983 &0.089 &\underline{0.262} &\textbf{0.268} &\textbf{198.240} \\
\slim &4401.2 &0.098 &\textbf{0.965} &0.089 &0.493 &0.063 &228.155 \\
MultiVAE &4249.6 &0.112 &0.900 &0.083 &0.426 &0.112 &255.457 \\
\itemknn &\textbf{7100.2} &\underline{0.188} &0.875 &0.079 &\textbf{0.272} &\underline{0.262} &\underline{199.488} \\
MF2020 &1631.2 &0.040 &0.766 &0.074 &0.776 &0.009 &320.867 \\
NeuMF &4647.0 &0.108 &0.756 &0.071 &0.542 &0.105 &276.817 \\
BPRMF &1793.6 &0.026 &0.693 &0.067 &0.640 &0.012 &409.832 \\
\ials &2052.0 &0.078 &0.730 &0.066 &0.654 &0.016 &202.718 \\
MostPop &72.0 &0.002 &0.244 &0.027 &1.000 &0.000 &629.745 \\
Random &8443.6 &0.875 &0.011 &0.001 &0.079 &0.738 &27.654 \\
\bottomrule
\end{tabular}
\end{minipage}
\end{table*} 

Generally, we observe that the ranking of the algorithms is not entirely consistent across the datasets. Here, we summarize a number of patterns that we observed, having in mind that beyond-accuracy measures are only of secondary interest in this study.

\begin{itemize}
  \item For \textbf{ARP}, which reports the average item popularity in the top-\emph{n} lists, we find that BPRMF often has the strongest tendency to recommend popular items on all datasets. MF2020 and \easer are also often at the higher end regarding the popularity bias. The ranking of the algorithms however varies across datasets. On the \ml dataset, the differences between algorithms are also generally smaller than for other datasets. On the other end of the spectrum, we observe that the neural methods NeuMF and MultiVAE sometimes succeed to include less popular items in the recommendation lists. \pbeta and \itemknn are similarly successful on the Epinions and \amzmusic in this respect. The \textbf{APLT} metric, which considers the popularity and coverage of long-tail items are negatively correlated with the \textbf{ARP} metric, 
      i.e., the more popular items are recommended, the fewer from the long tail.
  \item The novelty metrics \textbf{EPC} and \textbf{EFD}, like all remaining beyond-accuracy metrics considered here, are generally negatively correlated with the \textbf{ARP} metric as well. 
      An interesting pattern here is that models that perform well on the nDCG are also mostly highly ranked in terms of the novelty metrics.
  \item Looking at the fairness metric \textbf{PREO}, which is also based on item popularity and where lower values are better, the picture is not so clear. The neural MultiVAE method, for example, seems to rather consistently produce relatively fair recommendations according to this metric. \itemknn leads to very good results on the Epinions and Amazon dataset, and to average performance on the \ml dataset. For this latter dataset, the spread of values is however not too high.
  \item Finally, considering the \textbf{Gini} index, MultiVAE generally leads to lower concentration levels on \ml, and \itemknn and \pbeta have lower concentration effects for the Epinions and \amzmusic datasets.
      Looking at \textbf{Item Coverage}, both nearest-neighbor methods and the neural approaches are typically better than the matrix factorization techniques \ials and BPRMF. The patterns are however not consistent across datasets. \easer, for example, leads to relatively high item coverage on \amzmusic, but not on the other datasets.
\end{itemize}

Overall, not many consistent patterns regarding beyond-accuracy measures across all three datasets can be observed. One example of such a pattern is a certain popularity bias of the BPRMF method, which was previously observed~\cite{JannachLercheEtAl2015}. Some patterns, like good item coverage for \itemknn, are only found for the \amzmusic and Epinions datasets, which suggests that the widely used \ml dataset may be to some extent unique and it stands to question how representative this dense dataset is for other typical application scenarios, e.g., for e-commerce settings.


\subsection{Time Measurements}
We carried out all experiments on a computing cluster of our organization. The used cluster is based on IBM Power9 processors and has 980 nodes.
Each node is equipped with 32 cores and 4 NVIDIA Volta GPUs. One cluster node with 200GB  of RAM with 4 logical CPUs was reserved for each experiment. In addition, one NVIDIA Volta GPU with 16GB of RAM has been allocated for the experiments with the neural models NeuMF and MultiVAE.
Table~\ref{tab:timings-table} shows the time measurements obtained for the three datasets, using the optimal parameters (e.g., number of latent factors) that were determined through hyperparameter tuning. The numbers reported in the table refer to the time needed (in seconds) to train the model once, and to create and evaluate the recommendation lists for all users in the test set.

\color{black}
\begin{table*}[!htp]
\caption{Training and evaluation time}
\label{tab:training_time}
\begin{subtable}{0.33\textwidth}
\centering
\small
\renewcommand{\arraystretch}{0.8}
\rowcolors{1}{white}{gray!15}
\begin{tabularx}{0.66\textwidth}{Xr}\toprule
\textbf{Algorithm} &\textbf{time (sec.)} \\\midrule
MF2020 &\num{1.53e+4} \\
NeuMF &\num{7.97e+3} \\
BPRMF &\num{3.97e+3} \\
\ials &331.93 \\
\userknn &87.29 \\
\easer &85.93 \\
\slim &73.19 \\
MultiVAE &67.03 \\
\pbeta &47.06 \\
\itemknn &42.74 \\
Random &27.49 \\
MostPop &24.63 \\
\bottomrule
\end{tabularx}
\caption{MovieLens-1M}\label{tab:time-ml1m}
\end{subtable}
\begin{subtable}{0.33\textwidth}
\centering
\small
\renewcommand{\arraystretch}{0.8}
\rowcolors{1}{white}{gray!15}
\begin{tabularx}{0.66\textwidth}{Xr}\toprule
\textbf{Algorithm} &\textbf{time (sec.)} \\\midrule
NeuMF &\num{3.57e+4} \\
\ials &\num{2.90e+4} \\
MF2020 &\num{2.65e+4} \\
MultiVAE &\num{1.37e+4} \\
\easer &\num{1.85e+3} \\
BPRMF &\num{1.51e+3} \\
\slim &403.29 \\
\pbeta &270.78 \\
\itemknn &257.96 \\
\userknn &247.68 \\
Random &50.86 \\
MostPop &45.58 \\
\bottomrule
\end{tabularx}
\caption{Amazon Digital Music}\label{tab:time-amazon}
\end{subtable}
\begin{subtable}{0.33\textwidth}
\centering
\small
\renewcommand{\arraystretch}{0.8}
\rowcolors{1}{white}{gray!15}
\begin{tabularx}{0.66\textwidth}{Xr}\toprule
\textbf{Algorithm} &\textbf{time (sec.)} \\\midrule
\ials &\num{3.27e+4} \\
MF2020 &\num{1.97e+4} \\
NeuMF &\num{3.46e+3} \\
BPRMF &\num{2.26e+3} \\
\easer &\num{1.12e+3} \\
\slim &344.69 \\
MultiVAE &215.43 \\
\pbeta &148.24 \\
\userknn &144.05 \\
\itemknn &139.42 \\
Random &47.99 \\
MostPop &44.24 \\
\bottomrule
\end{tabularx}
\caption{Epinions}\label{tab:time-epinions}
\end{subtable}
\label{tab:timings-table}
\end{table*}

The results show that there is a substantial spread between the algorithms. While there are some models that complete
training and testing 
within one minute, training the MF2020 method on the ML1M dataset, where it performed well, took several days.
We note here that more efficient implementations of matrix factorization techniques have been proposed~\cite{DBLP:journals/corr/abs-2110-14044}. 
Also the NeuMF model needed substantial time to complete the computations. In contrast, the MultiVAE model, which was also originally evaluated on larger datasets in~\citet{DBLP:conf/www/LiangKHJ18} was among the fastest models. The neighborhood-based models and \pbeta were also implemented for high efficiency. For the other datasets, Epinions and Amazon, the results are similar with NeuMF and the matrix factorization models often taking substantial computation time. For this latter class of models, the efficiency also largely depends on the optimal number of latent factors. 

Generally, combining the timing results with accuracy results from above, we see no clear indication for the given datasets that computationally more complex models are favorable in terms of prediction accuracy.


\section{Summary, Discussion \& Outlook}\label{sec:summary} 
In recent years, several researchers have identified major challenges with respect to reproducibility and progress in recommender systems research. Various factors contribute to these phenomena, in particular \emph{(a)} that a larger fraction of published research is not reproducible because authors do not share the required artifacts and \emph{(b)} that the experiments in published research mainly aim to highlight the superiority of a new model. In the context of this latter aspect, this practically often means that only the new method is carefully fine-tuned but not the compared baseline methods. Furthermore, the choice of the baselines is sometimes limited to very recent models, thus probably missing strong baselines that were published earlier.

With this work, our goal is to address these open issues in different ways. First, we conducted a large number of reproducible experiments on different datasets and involving a variety of algorithms from different families in order to provide an independent evaluation of existing techniques along different quality and performance measures. The outcomes of these experiments shall help guide researchers in the choice of baseline algorithms to consider in their own research. In particular we found that one should consider algorithms of different types in any evaluation, as there appears to be no single method that is better than all others in all experimental configurations. Second, we ran these experiments with the help of a recent general evaluation framework for recommender systems \cite{DBLP:conf/sigir/AnelliBFMMPDN21}, 
thus allowing other researchers to benchmark their new models within a defined environment and against already well-tuned baselines.

In terms of the outcomes of the experiments, our reproducibility study confirmed earlier findings that the latest models are not often the best performing ones, in particular for the modest-sized datasets that we considered in our evaluation. In our ongoing and future work, we plan to fine-tune our models also on larger datasets and to share these tuned models publicly. Thereby, we hope to reduce the often huge computational effort that other researchers would otherwise need to fine-tune all baseline models whenever they propose a new model. Over time, this collection of fine-tuned models for various datasets may represent a step towards a shared understanding of what represents the ``state-of-the-art'' in algorithms research. For these larger datasets, we also expect a more consistent and strong performance of deep learning models.

Besides accuracy metrics, our experiments included a number of beyond-accuracy metrics relating to popularity bias, novelty, fairness, and item coverage. Our results confirm earlier findings that there can be substantial differences between algorithms, e.g., in terms of their tendency to recommend popular items. Such algorithm tendencies can be of high relevance in practical application settings, e.g., when the goal is to support item discovery through the recommendations. An important observation in our research is that common metrics for novelty and fairness are tightly coupled and correlated with general popularity biases\footnote{In theory, the Gini index is not necessarily tied to popularity biases, but with the typical long-tail distributions it usually captures a concentration of items in the ``short head''.}. Future research might therefore strive to find alternative metrics that more often go beyond popularity as indicators for novelty, diversity, fairness, or serendipity.

In addition to this, a careful analysis on the effect of the optimization goals for hyperparameter tuning is missing in the literature. The results presented herein considered methods optimized for one specific accuracy-oriented metric, i.e., nDCG. But what would happen if other metrics are used for this optimization? It is true that there are strong correlations between some metrics, as discussed before, but it is also well-known that accuracy and beyond-accuracy measurements are typically inversely related, hence, the question of what ``state-of-the-art'' means in terms of these other metrics remains open and should be addressed in the future.

Finally, another aspect regarding the splitting strategy has to be taken into consideration. Here, we adopted a random hold-out splitting strategy with repeated experiments that became popular in recent literature. Together with k-folds cross-validation, they are representative of the evaluation protocols adopted in recent works. Nevertheless, random-based splitting strategies undoubtedly favor some methods since information regarding the future general users' behavior is exploited in the training phase. More realistic time-aware splitting strategies should be investigated to study how much they impact the overall ranking of recommendation systems.


\begin{acks}
We acknowledge the CINECA award under the ISCRA initiative, for the availability of high performance computing resources and support.
The authors acknowledge partial support of the projects: PID2019-108965GB-I00, Fincons CdP3, PASSPARTOUT, Servizi Locali 2.0, ERP4.0, Secure Safe Apulia.
\end{acks}


\bibliographystyle{ACM-Reference-Format}
\bibliography{references}

\appendix
\section*{APPENDIX}
\section{Inter-Metric Correlations - Beyond Accuracy Metrics}
In this section we report the correlations between each pair of beyond-accuracy metrics. For each dataset, the tables indicate the Pearson product-moment correlation coefficient, unveiling strong direct and inverse correlations. Please note we do not report same analysis for accuracy metrics here, since the topic of correlation among those metrics has been extensively studied in prior literature. Please refer to~\citet{DBLP:conf/recsys/ValcarceBPC18}, and~\citet{DBLP:conf/recsys/AnelliNSPR19} for further details.
\begin{table}[h]
\caption{Detailed Metric Correlations. The tables show how much each beyond-accuracy metric (computed on recommendation lists of ten items for each user) correlates with each other. Specifically, the table shows the Pearson product-moment correlation coefficient for each dataset.}
\footnotesize

\rowcolors{1}{white}{gray!15}
\begin{tabular}{lcccccccc}
\hline
Movielens & \textbf{EFD} & \textbf{Gini} & \textbf{IC} & \textbf{PopREO} & \textbf{PopRSP} & \textbf{ACLT} & \textbf{APLT} & \textbf{ARP} \\ \hline
\textbf{EPC} & 1.00 & -0.73 & -0.36 & 0.75 & 0.79 & -0.79 & -0.79 & 0.20 \\
\textbf{EFD} &  & -0.74 & -0.37 & 0.77 & 0.81 & -0.80 & -0.80 & 0.23 \\
\textbf{Gini} &  &  & 0.86 & -0.99 & -0.99 & 0.99 & 0.99 & -0.81 \\
\textbf{IC} &  &  &  & -0.86 & -0.79 & 0.80 & 0.80 & -0.95 \\
\textbf{PopREO} &  &  &  &  & 0.99 & -0.99 & -0.99 & 0.78 \\
\textbf{PopRSP} &  &  &  &  &  & -1.00 & -1.00 & 0.74 \\
\textbf{ACLT} &  &  &  &  &  &  & 1.00 & -0.74 \\
\textbf{APLT} &  &  &  &  &  &  &  & -0.74 \\
\end{tabular}

\rowcolors{1}{white}{gray!15}
\begin{tabular}{lcccccccc}
\hline
Amazon & \textbf{EFD} & \textbf{Gini} & \textbf{IC} & \textbf{PopREO} & \textbf{PopRSP} & \textbf{ACLT} & \textbf{APLT} & \textbf{ARP} \\ \hline
\textbf{EPC} & 1.00 & -0.10 & 0.45 & -0.26 & 0.09 & -0.04 & -0.04 & -0.46 \\
\textbf{EFD} &  & -0.05 & 0.49 & -0.32 & 0.03 & 0.02 & 0.02 & -0.50 \\
\textbf{Gini} &  &  & 0.78 & -0.85 & -0.96 & 0.88 & 0.88 & -0.69 \\
\textbf{IC} &  &  &  & -0.93 & -0.74 & 0.70 & 0.70 & -0.88 \\
\textbf{PopREO} &  &  &  &  & 0.87 & -0.87 & -0.87 & 0.84 \\
\textbf{PopRSP} &  &  &  &  &  & -0.97 & -0.97 & 0.60 \\
\textbf{ACLT} &  &  &  &  &  &  & 1.00 & -0.58 \\
\textbf{APLT} &  &  &  &  &  &  &  & -0.58 \\
\end{tabular}

\rowcolors{1}{white}{gray!15}
\begin{tabular}{lcccccccc}
\hline
Epinions & \textbf{EFD} & \textbf{Gini} & \textbf{IC} & \textbf{PopREO} & \textbf{PopRSP} & \textbf{ACLT} & \textbf{APLT} & \textbf{ARP} \\ \hline
\textbf{EPC} & 0.83 & -0.05 & -0.11 & -0.94 & -0.84 & 0.91 & 0.91 & -0.81 \\
\textbf{EFD} &  & -0.54 & -0.59 & -0.62 & -1.00 & 0.97 & 0.97 & -0.67 \\
\textbf{Gini} &  &  & 1.00 & -0.25 & 0.55 & -0.43 & -0.43 & -0.18 \\
\textbf{IC} &  &  &  & -0.19 & 0.60 & -0.49 & -0.49 & -0.12 \\
\textbf{PopREO} &  &  &  &  & 0.63 & -0.74 & -0.74 & 0.81 \\
\textbf{PopRSP} &  &  &  &  &  & -0.99 & -0.99 & 0.64 \\
\textbf{ACLT} &  &  &  &  &  &  & 1.00 & -0.69 \\
\textbf{APLT} &  &  &  &  &  &  &  & -0.69 \\\hline
\end{tabular}

\label{tab:dataset_metric_correlations}
\end{table}

\section{F1 scores - Additional numerical examples}

The F1 score 
represents the harmonic mean of Precision and Recall. 
In the recommendation domain, when evaluating lists of k items (top-k evaluation), it is usually defined as follows:
\begin{equation}\label{eq:puf1}
    F1\ Score = \frac{1}{|U|}\sum\limits_{u \in U}{2 * \frac{P_u@k * R_u@k}{P_u@k + R_u@k}}
\end{equation}
where $U$ is the set of the users in the population, and where $Pu@k$ and $Ru@k$ are the Precision and Recall values for a user u's \emph{top-k} recommendations, respectively. In an alternative formulation, the F1 Score could be computed \emph{after} obtaining the average Precision and Recall values across all users:

\begin{align}
    P@k &= \frac{1}{|U|}\sum\limits_{u \in U}{P_u@k} \\
    R@k &= \frac{1}{|U|}\sum\limits_{u \in U}{R_u@k}\\
    F1\ Score &= 2 * \frac{P@k * R@k}{P@k + R@k\label{eq:avgf1}}
\end{align}

These alternative formulations may lead to different results, as we highlight in the following examples.
Let us consider a population of five users for whom we have computed the Precision and Recall values for a recommendation system A (see Table~\ref{tab:example1}).

\begin{table}[!h]\centering
\caption{Accuracy results for the toy recommendation systems. P@k, R@k, and F@k stands for individual Precision, Recall, and F1 Score with a list of $k$ recommendations, respectively. \emph{Average} reports the overall Precision and Recall values. Per-user F1 and average-based F1 indicates the F1 scores computed using Equation~\ref{eq:puf1} and Equation ~\ref{eq:avgf1}, respectively.}
\footnotesize
\begin{subtable}{0.48\textwidth}
\begin{table}[H]

\small
\rowcolors{2}{white}{gray!15}
\begin{tabular}{lrrrr}\toprule
Population &\textbf{$P_u@k$} &\textbf{$R_u@k$} &\textbf{$F_u@k$} \\\midrule
\textbf{$user_0$} &0.2 &0.3 &0.240 \\
\textbf{$user_1$} &0.5 &0.6 &0.545 \\
\textbf{$user_2$} &0.3 &0.4 &0.343 \\
\textbf{$user_3$} &0.6 &0.3 &0.400 \\
\textbf{$user_4$} &0.2 &0.3 &0.240 \\ \midrule
&\textbf{$P@k$} &\textbf{$R@k$} &\textbf{$F@k$} \\
\textbf{Average} &0.36 &0.38 & \\
\textbf{Per-user F1} & & &0.354 \\
\textbf{Average-based F1} & & &0.370 \\
\bottomrule
\end{tabular}
\caption{Toy recommendation system A.}\label{tab:example1}
\end{table}
\end{subtable}
\begin{subtable}{0.48\textwidth}
\begin{table}[H]

\small
\rowcolors{2}{white}{gray!15}
\begin{tabular}{lrrrr}\toprule
Population &\textbf{$P_u@k$} &\textbf{$R_u@k$} &\textbf{$F_u@k$} \\\midrule
\textbf{$user_0$} &0.2 &0.4 &0.267 \\
\textbf{$user_1$} &0.5 &0.2 &0.286 \\
\textbf{$user_2$} &0.4 &0.4 &0.400 \\
\textbf{$user_3$} &0.2 &0.6 &0.300 \\
\textbf{$user_4$} &0.5 &0.4 &0.444 \\ \midrule
&\textbf{$P@k$} &\textbf{$R@k$} &\textbf{$F@k$} \\
\textbf{Average} &0.36 &0.40 & \\
\textbf{Per-user F1} & & &0.339 \\
\textbf{Average-based F1} & & &0.379 \\
\bottomrule
\end{tabular}
\caption{Toy recommendation system B.}\label{tab:example2}
\end{table}
\end{subtable}
\end{table}

It is worth noticing that the F1 formulation from Equation~\ref{eq:puf1}, denoted as \emph{Per-User F1}, returns an F1 score that is lower than the overall averaged values of Precision and Recall. This can happen due to the product of individual Precision and Recall values. If one of the two is small, it affects the result and impacts the F1 score. Conversely, this behavior is not likely to occur when the F1 is computed on already averaged Precision and Recall values (Average-based F1).


Furthermore, suppose that we evaluate the performance of two recommender systems, A and B (Table~\ref{tab:example2}).
The two systems lead to the same average Precision value, and B leads to a higher Recall value than A.
It may now be surprising to see that A has a higher \emph{per-user} F1 score than B. As a consequence of the previously discussed phenomenon, it is indeed possible. That is, although the Precision value of system B is equal to system A, some individual Precision values lead to poor individual F1 results that affect the overall value of the metric. Some  examples of such cases can be found in the accuracy results of the paper.

\section{Hyperparameters Range}
\begin{table}[h!]\centering
\caption{Hyperparameter values for our baselines.}\label{tab:parmas}
\begin{tabular}{p{0.1\linewidth}p{0.15\linewidth}|wc{0.3\linewidth}wc{0.1\linewidth}wc{0.1\linewidth}}\toprule
Algorithm &Hyperparameter &Range &Type &Distribution \\\midrule
\multirow{2}{*}{\makecell{UserKNN,\\ItemKNN}} &topK &5 - 1000 &Integer &uniform \\
&similarity &\makecell{cosine, jaccard, dice,\\pearson, euclidean}&Categorical & \\\hline
\multirow{4}{*}{\pbeta} &topK &5 - 1000 &Integer &uniform \\
&alpha &0 - 2 &Real &uniform \\
&beta &0 - 2 &Real &uniform \\
&normalization &True, False &Categorical & \\\hline
\multirow{3}{*}{\slim} &topK &5 -1000 &Integer &uniform \\
&l1 ratio &0.00001 - 1 &Real &log-uniform \\
&alpha &0.01 - 1 &Real &uniform \\\hline
\easer &l2 norm &1 - 10000000 &Real &log-uniform \\\hline
\multirow{5}{*}{MF2020} &num factors &8, 16, 32, 64, 128, 256 &Integer & \\
&epochs &30 - 100 &Integer &uniform \\
&learning rate &0.00001 - 1 &Real &log-uniform \\
&reg &0.00001 - 0.1 &Real &log-uniform \\
&negative sample &4,6,8 &Integer & \\\hline
\multirow{5}{*}{\ials} &num factors &1 - 200 &Integer &uniform \\
&scaling &linear, log &Categorical & \\
&alpha &0.001 - 50 &Real &uniform \\
&epsilon &0.001 - 10 &Real &uniform \\
&reg &0.001 - 0.01 &Real &uniform \\\hline
\multirow{6}{*}{BPRMF} &num factors &8, 16, 32, 64, 128, 256 &Integer & \\
&learning rate &0.00001 - 1 &Real &log-uniform \\
&batch size &128, 256, 512 &Integer & \\
&reg user &0.00001 - 0.1 &Real &log-uniform \\
&reg positive item &0.00001 - 0.1 &Real &log-uniform \\
&reg negative item &0.00001 - 0.1 &Real &log-uniform \\\hline
\multirow{5}{*}{NeuMF} &num factors &8, 16, 32, 64, 128, 256 &Integer & \\
&epochs &30 - 100 &Integer &uniform \\
&learning rate &0.00001 - 1 &Real &log-uniform \\
&batch size &128, 256, 512 &Integer & \\
&negative sample &4,6,8 &Integer & \\\hline
\multirow{6}{*}{MultiVAE} &epochs &100 - 300 &Integer &uniform \\
&learning rate &0.00001 - 1 &Real &log-uniform \\
&batch\_size &64, 128, 256 &Integer & \\
&intermediate dim &400 - 800 &Integer &uniform \\
&latent dim &100-400 &Integer &uniform \\
&reg &0.00001 - 1 &Real &log-uniform \\
\bottomrule
\end{tabular}
\end{table}

\begin{table}[!htp]\centering
\caption{Hyperparameter values for our baselines on all datasets.}\label{tab:params_datasets}
\begin{tabular}{p{0.1\linewidth}p{0.15\linewidth}|wc{0.15\linewidth}wc{0.15\linewidth}wc{0.15\linewidth}}\toprule
Algorithm &Hyperparameter &MovieLens &Amazon &Epinions \\\midrule
\multirow{2}{*}{UserKNN} &topK &117 &226 &139 \\
&similarity &correlation &cosine &cosine \\\hline
\multirow{2}{*}{ItemNN} &topK &95 &798 &137 \\
&similarity &cosine &cosine &cosine \\\hline
\multirow{4}{*}{\pbeta} &topK &158 &803 &144 \\
&alpha &1.4350197 &0.4973207 &0.8719344 \\
&beta &0.3265517 &0.2836938 &0.2483698 \\
&normalization &true &false &true \\\hline
\multirow{3}{*}{\slim} &topK &518 &663 &663 \\
&l1 ratio &0.0000420 &0.0000108 &0.0000108 \\
&alpha &0.2978543 &0.0486771 &0.0486771 \\\hline
\easer &l2 norm &238.5621338 &238.5621338 &238.5621338 \\\hline
\multirow{5}{*}{MF2020} &num factors &128 &64 &16 \\
&epochs &72 &92 &97 \\
&learning rate &0.1295965 &0.1295965 &0.0154435 \\
&reg &0.0087583 &0.0125009 &0.0223642 \\
&negative sample &4 &8 &4 \\\hline
\multirow{6}{*}{\ials} &num factors &51 &200 &178 \\
&epochs &27 &70 &145 \\
&scaling &log &log &log \\
&alpha &6.3818930 &9.1219718 &2.8537184 \\
&epsilon &5.6496278 &0.4921936 &2.3098481 \\
&reg &0.0494734 &0.4921936 &0.0411491 \\\hline
\multirow{7}{*}{BPRMF} &num factors &256 &64 &256 \\
&epochs &73 &86 &63 \\
&learning rate &0.0378936 &0.1265624 &0.1004075 \\
&batch size &256 &256 &256 \\
&reg user &0.0157839 &0.0058673 &0.0002613 \\
&reg positive item &0.0005651 &0.0052985 &0.0034511 \\
&reg negative item &0.0012779 &0.0009577 &0.0328127 \\\hline
\multirow{5}{*}{NeuMF} &num factors &16 &128 &32 \\
&epochs &93 &100 &39 \\
&learning rate &0.0000366 &0.0001365 &0.0000465 \\
&batch size &256 &64 &256 \\
&negative sample &6 &6 &8 \\\hline
\multirow{6}{*}{MultiVAE} &epochs &100 &205 &200 \\
&learning rate &0.0001545 &0.0000723 &0.0001003 \\
&batch\_size &128 &128 &128 \\
&intermediate dim &674 &721 &674 \\
&latent dim &175 &279 &175 \\
&reg &0.0000105 &0.1153400 &0.0020018 \\
\bottomrule
\end{tabular}
\end{table}

\end{document}